\documentclass{rmf-d}
\usepackage{nopageno,rmfbib,multicol,times,epsf,amsmath,amssymb,cite}
\usepackage[latin1]{inputenc}
\usepackage[]{caption2}
\usepackage{graphics}
\usepackage[demo]{graphicx}
\usepackage{hyperref}
\usepackage{wrapfig, blindtext}
\usepackage{comment}
\usepackage{tabularx}
\usepackage{float}
\def\rmfcornisa{Research OR Education of Physics \hfill\rmf\ {\bf ??} (*?*) ???--???
\hfill MES? A\~NO?}

\def\rmfcintilla{{\it Rev.\ Mex.\ Fis.\/} {\bf ??} (*?*) (????) ???--???}
\clearpage \rmfcaptionstyle 
\begin{document}
\title{Chiral condensates for neutron stars in hadron-quark crossover; \\
from a parity doublet nucleon model to an NJL quark model 
\vspace{-6pt}}

\author{Takuya Minamikawa}
\address{Department of Physics, Nagoya University, Nagoya 464-8602, Japan}

\author{Toru Kojo}
\address{Key Laboratory of Quark and Lepton Physics (MOE) and Institute of Particle Physics,
Central China Normal University, Wuhan 430079, China}

\author{Masayasu Harada}
\address{Department of Physics, Nagoya University, Nagoya 464-8602, Japan}
\address{Kobayashi-Maskawa Institute for the Origin of Particles and the Universe, Nagoya University, Nagoya 464-8602, Japan}
\address{Advanced Science Research Center, Japan Atomic Energy Agency, Tokai 319-1195, Japan}

\maketitle

\recibido{day month year}{day month year\vspace{-12pt}}

\begin{abstract}
\vspace{1em}
In this contribution, 
we summarize our recent studies on 
the chiral invariant mass and the chiral condensates in neutron star matter.
We construct a unified equations of state 
assuming the crossover phase transition 
from hadronic matter described by a parity doublet model 
to quark matter by an Nambu--Jona-Lasinio type quark model.
We first show that 
the chiral invariant mass is constrained to be 
$600\,\mbox{MeV}\, \lesssim\, m_0 \,\lesssim \, 900\,\mbox{MeV}$ 
from recent observations of neutron stars.
We then determine the density dependence of the chiral condensate in the crossover description, 
and show that the chiral condensates are actually smoothly connected 
from the hadronic matter where the change is driven by the positive chiral scalar charge in a nucleon, 
to the quark matter where the change is by the modification of the quark Dirac sea, 
reflecting the hadron-quark crossover.
\vspace{1em}
\end{abstract}

\keys{
Chiral symmetry, 
Chiral condensate, 
Parity doublet model, 
High density nuclear matter, 
Neutron star, 
\vspace{-4pt}}
\pacs{   \bf{\textit{PLEASE PROVIDE }}    \vspace{-4pt}}
\begin{multicols}{2}

%PLACE THE TEXT HERE 
%* Author's note:
%All the macros have to be in the same folder as the main tex file

%%%%%%%%%%%%%%%%%%%%%%%%%%%%%%%%%%%%%%%%%%%%%%%%%%%%%%%%%%%%%%%%
%	INTRODUCTION
%%%%%%%%%%%%%%%%%%%%%%%%%%%%%%%%%%%%%%%%%%%%%%%%%%%%%%%%%%%%%%%%

This talk is mainly based on our papers of Refs.\cite{Minamikawa:2020jfj,Minamikawa:2021fln}. 

\section{Introduction}

Chiral symmetry is a symmetry of quantum chromodynamics (QCD) which 
plays an important role in the hadron dynamics at low energy.
In vacuum the chiral symmetry is spontaneously broken by the condensation of quark-antiquark pairs, 
called chiral condensate $\sigma$. 
The change of vacuum structure produces the quark mass gap of $\sim 300$ MeV.
In high energy region, such as in high temperature or in high density environment, 
the chiral symmetry is restored with vanishing $\sigma$.
In a class of effective models, the majority of the nucleon mass comes from the chiral condensate. 
On the other hand, the lattice simulation in Ref.~\cite{Aarts:2017rrl} shows that 
the nucleon mass remains large even at a temperature with substantial reduction of $\sigma$.
One of the effective models which reflect this property is the parity doublet model 
\cite{Detar:1988kn,Jido:2001nt}
in which nucleons have substantial chiral invariant masses.

One of laboratories to study the roles of the chiral symmetry is a neutron star 
in which a matter at supra nuclear densities are realized. 
The density of a neutron star reaches baryon densities ($n_B$) several times greater than 
the nuclear saturation density $n_0\approx0.16\,\mathrm{fm}^{-3}$, 
and may accommodate quark matter at the core.
The transition from hadronic to quark matter, 
which is supposed to occur in neutron stars, should give us 
insight into the nature of chiral symmetry breaking and its restoration.

One of the key observables in neutron star physics is mass-radius sequence of neutron stars ($M$-$R$ relation), 
which has a one-to-one correspondence with the underlying equation of state (EOS).
The shape of the $M$-$R$ curve is correlated with 
the stiffness of EOS at several fiducial densities \cite{Kojo:2019raj}. 
The low density part ($n_B\lesssim 2n_0$) is largely correlated with the overall radii of neutron stars 
while the high density part ($n_B \gtrsim 3$-$5n_0$) determines the maximum mass.
The neutron star observations and nuclear physics suggest 
EOS should be soft enough to pass the constraints from the tidal deformability in GW170817 event, 
$R_{1.4} \lesssim 13$ km for 1.4$M_\odot$ neutron star.
Meanwhile the heaviest neutron star currently known is 
PSR J0740+6620 with the mass $2.08\pm 0.07M_\odot$ and the radius $\simeq 12.4 \pm 0.7$ km, 
requiring EOS to be very stiff at high density.
This soft-to-stiff combination disfavors 
radical softening and strong first order phase transitions for $\sim 2$-$5n_0$. 
This leads us to the hadron-quark continuity picture 
as a baseline to describe EOS from hadronic to quark matter. 
The possible weak first order phase transitions may be treated as perturbations.
In practice we consider the 3-window modeling~\cite{Baym:2017whm}: 
we use a parity doublet model (PDM) for $n_B \le 2n_0$, 
a Nambu--Jona-Lasinio (NJL) type quark model for $n_B \ge 5n_0$, 
and interpolate them to describe EOS at $2$-$5n_0$.
Our interpolation is highly constrained by neutron star observables and the causality condition with the sound velocity less than the light velocity.

The comparison of our $M$-$R$ relations with observations give us 
the estimates on the range of our model parameters and microscopic insights into the dynamics.
Actually, one can proceed further by extending the interpolation from physical EOS, 
$P(\mu_B)$, to a general generating function, $P(\mu_B; {\bf J})$, 
in which external sources ${\bf J}$ are coupled to various condensates and flavor densities.
After constructing a unified functional with various sources, 
one can differentiate the functional with ${\bf J}$ to compute various quantities 
such as chiral and diquark condensates as well as flavor compositions \cite{Minamikawa:2021fln}.
This updates the 3-window modeling from thermodynamic quantities to more microscopic quantities 
which are not directly measured but are important to characterize the state of matter in dense QCD.

%%%%%%%%%%%%%%%%%%%%%%%%%%%%%%%%%%%%%%%%%%%%%%%%%%%%%%%%%%%%%%%%
%	MODELS AND RESULTS
%%%%%%%%%%%%%%%%%%%%%%%%%%%%%%%%%%%%%%%%%%%%%%%%%%%%%%%%%%%%%%%%

\section{Models }

In this section, we explain a PDM for hadronic matter
and an NJL model for quark matter.

\subsection{Parity doublet model for hadronic matter}

A PDM contains two iso-doublets of nucleon fields $\psi_{1,2}$ 
to describe positive and negative parity nucleons.
The fields $\psi_1$ and $\psi_2$ acquire the opposite phases under chiral transformations, 
\begin{align}
\psi_{1L}\to g_L\psi_{1L}\,,\quad\psi_{1R}\to g_R\psi_{1R}\,,\\
\psi_{2L}\to g_R\psi_{2L}\,,\quad\psi_{2R}\to g_L\psi_{2R}\,, 
\end{align}
where $g_L$ and $g_R$ are elements of SU(2)$_L$ and SU(2)$_R$ chiral groups respectively, 
and left-handed and right-handed fields are defined as 
\begin{align}
\psi_{iL}=\frac{1-\gamma_5}{2}\psi_i\,,\quad
\psi_{iR}=\frac{1+\gamma_5}{2}\psi_i\,, 
\end{align}
for $i=1,2$. 

Such assignment of the chirality is called mirror assignment.
With these fields, one can construct terms including a chiral invariant mass $m_0$, as 
\begin{align}
m_0(\bar\psi_1\gamma_5\psi_2+h.c.)\,. 
\end{align}
There are also Yukawa terms as in ordinary linear sigma models, $ g_{1} \bar{\psi}_{1} \sigma \psi_{1} $ and $ g_{2} \bar{\psi}_{2} \sigma \psi_{2} $. 
The effective potential of $\sigma$ field is minimized for a non-zero $\sigma$, 
which produces the chiral variant mass for nucleons. 
The diagonalization of chiral invariant and variant components leads to the mass spectra
\begin{align}\label{eq-nucleon-mass}
m_{N, N^\ast } = \sqrt{m_0^2+G_+^2\sigma^2}\mp G_-\sigma
\end{align}
where $G_\pm = |g_1\pm g_2|/2$. 
For vanishing chiral condensate, the nucleon remains massive; $m_N (m_{N^\ast})\to m_0$ as $\sigma\to0$. 
We note that, for larger $m_0$, the mass of nucleon is dominated by the chiral invariant mass $m_0$ 
so that $G_\pm$ should be smaller. 
Namely, the nucleon fields $\psi_{1,2}$ couples to $\sigma$ field weakly for a large $m_0$.

\begin{figure*}\centering
\includegraphics[width=0.4\hsize]{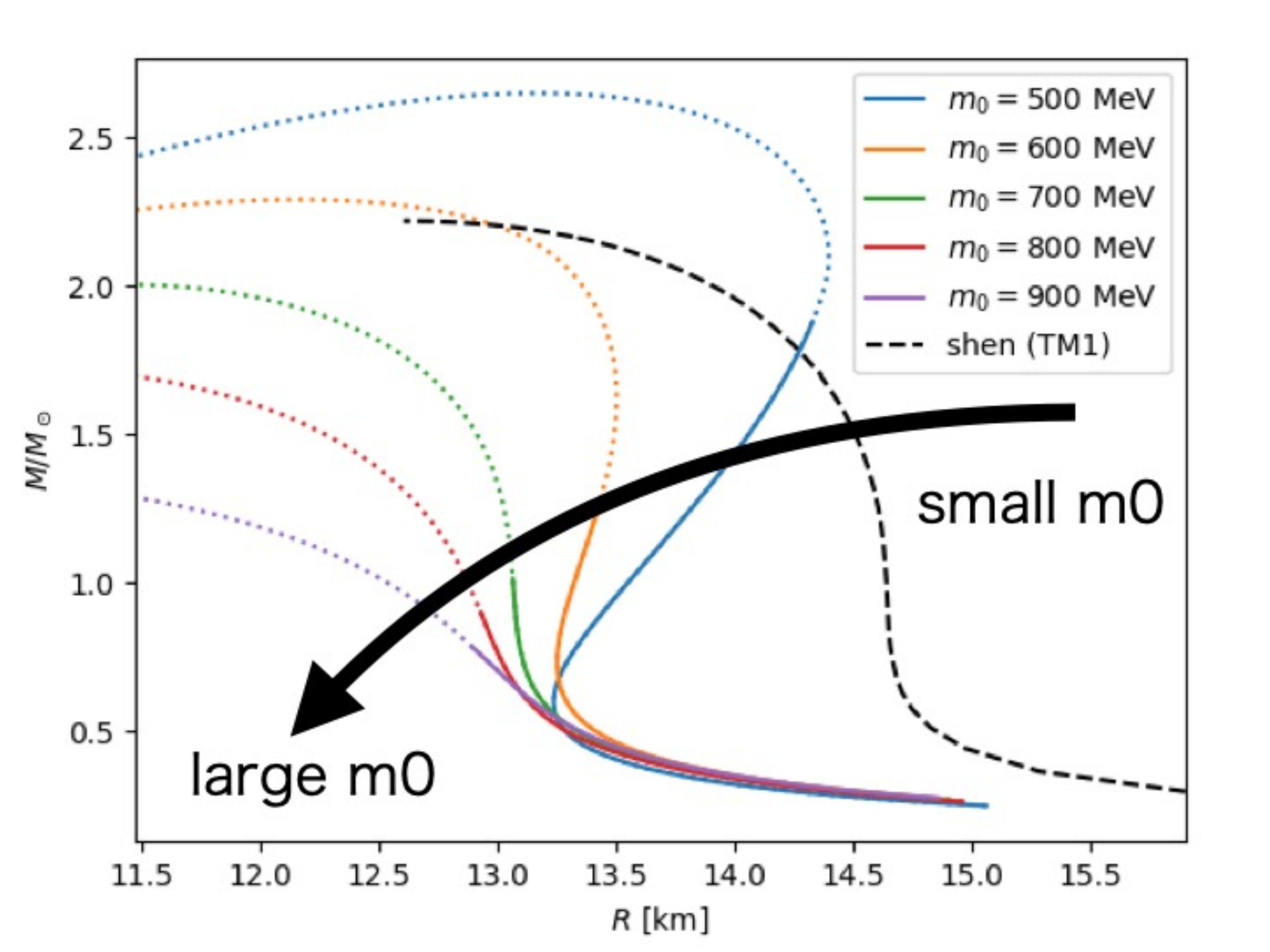}
\caption{
$M$-$R$ relations obtained from 
the PDM for $m_0=500,600,700,800,$ and $900\,\mathrm{MeV}$. 
The solid curves show the $M$-$R$ relations for $n_c<2n_0$, where $n_c$ is the central density. }
\label{fig-PDM_MR_for_m0}
\end{figure*}

For descriptions of hadronic matter, in addition we include $\omega$ and $\rho$ meson fields, 
which are necessary to reproduce the nuclear saturation properties at $n_B=n_0$ and the symmetry energy. 
For symmetric nuclear matter, the attractive contribution from $\sigma$ field 
and repulsive contribution from $\omega$ field are balanced. 
We were able to tune parameters to reproduce the saturation properties, 
but the high density extrapolation is very sensitive to the values of $m_0$. 
For large $m_0$, the $N$-$\sigma$ coupling is weak. 
Then, the $N$-$\omega$ coupling needed to reproduce 
the saturation properties is small enough to be balanced with the $\sigma$ attraction. 
As density exceeds $n_0$, weaker $\omega$ repulsion leads to softer EOS and smaller radius of neutron star. 
The trend is opposite for a small $m_0$.
The trend can be seen in Fig.\ref{fig-PDM_MR_for_m0}.

\subsection{NJL model for quark matter}

For quark matter we use the standard three-flavor NJL model with additional two terms: 
the interaction leading to the diquark condensates 
$H(q^t\Gamma_aq)(\bar{q}\Gamma^a\bar{q}^t)$
and the vector-type interaction $g_V(\bar{q}\gamma^{\mu}q)^2$. 
We explore the range of the coupling constants $(g_V, H)$ 
which are consistent with neutron star observations and the causality condition.

\subsection{Unified EOS with interpolation}

To construct a unified EOS, 
the EOSs of the PDM and the NJL model are interpolated with the following polynomial 
\begin{align}
P_\mathrm{Interp}=\sum_{i=0}^5a_i\mu_B^i \,.
\end{align}
The six coefficients $a_0,\cdots, a_5$ are determined by matching the polynomial 
with hadronic and quark EOS at the boundaries up to the second orders in derivatives.
The hadronic and quark boundaries are $n_B=2n_0$ and $5n_0$, respectively. 
The EOSs satisfying the causality condition $c_s^2=dP/d\varepsilon\leq1$ can be adopted as physical EOSs.

\section{Results}

\subsection{$M$-$R$ relation}

\begin{figure*}\centering
\includegraphics[width=0.4\hsize]{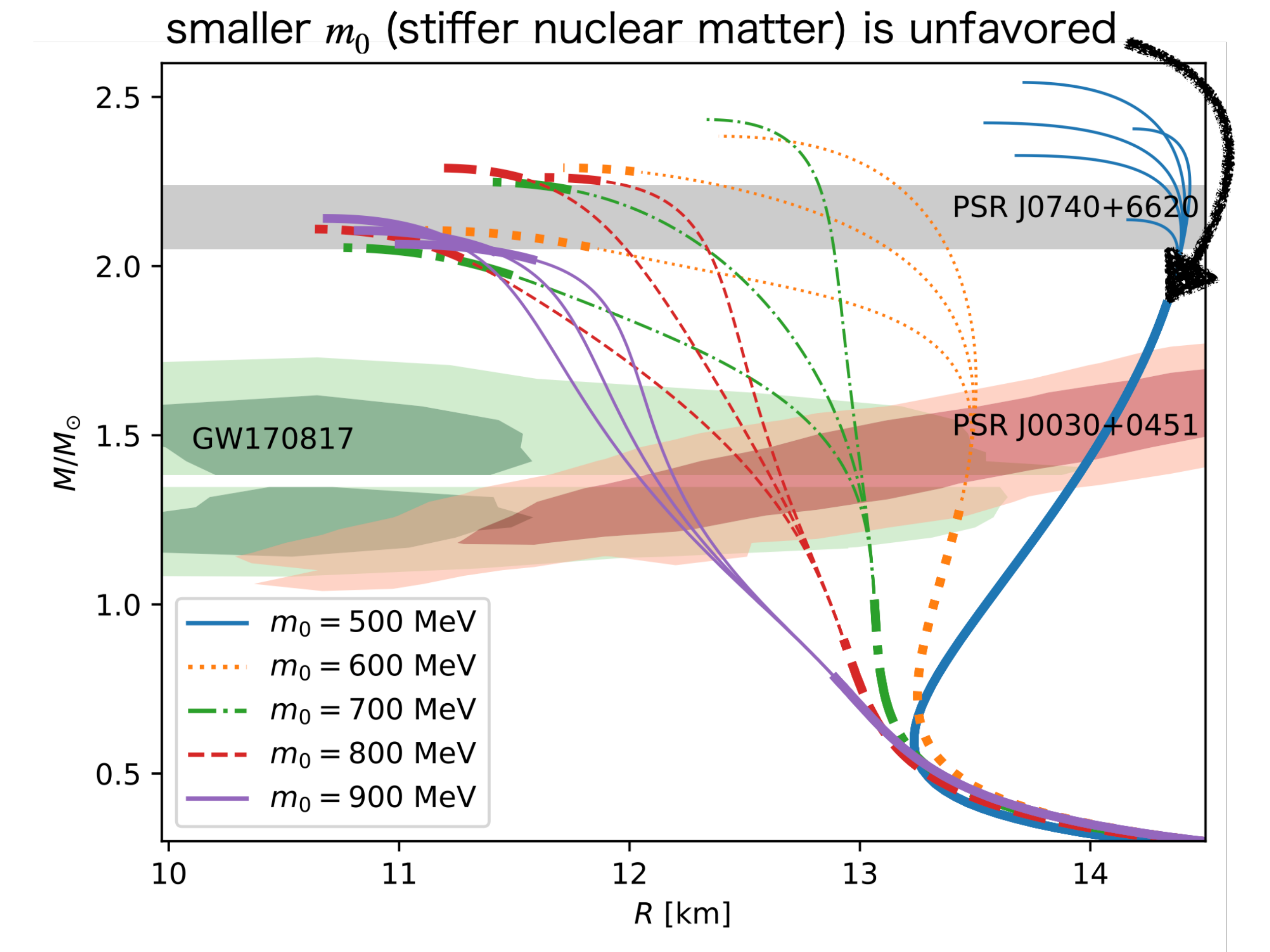}
\caption{
$M$-$R$ relations calculated from the unified EOSs. 
The solid blue curves pointed by the arrow are for $m_0=500\,\mathrm{MeV}$. }
\label{fig-unified_MR_all}
\end{figure*}

Shown in Fig.\ref{fig-unified_MR_all} are some selected results of $M$-$R$ relations for unified EOSs for various $m_0$~\cite{Minamikawa:2020jfj}. 
For neutron stars with the mass $\lesssim 0.5M_\odot$, the curves for different $m_0$ show roughly the same behavior. 
The bold parts of the curves correspond to EOS at $n_B \le 2n_0$, from which one can see that the overall radii of neutron stars are largely determined by hadronic EOS.

We note that, in Fig.~\ref{fig-unified_MR_all}, we show $M$-$R$ curves for which 
the maximum masses exceed the highest mass of the observed neutron star. 
The solid blue curves pointed by the arrow are for $m_0=500\,\mathrm{MeV}$, 
which correspond to the stiffest hadronic matter in our model. 
The radii are too large to be compatible with the constraints from GW170817 (the green shades).
Therefore, we omit the $m_0$ case and conclude that the favored range of $m_0$  is 
\begin{align}\label{eq-m0_constr}
600\,\mathrm{MeV}\lesssim m_0\lesssim900\,\mathrm{MeV}\,.
\end{align}
That implies that the large portion of the nucleon mass comes from the chiral invariant mass, 
and the $N$-$\sigma$ and $N$-$\omega$ couplings are 
weaker than the conventional linear $\sigma$ model with $m_0=0$.

\subsection{chiral condensate }

Our interpolation procedures for thermodynamic potentials can be extended to generating functionals with external sources ${\bf J}$.
We can estimate condensates coupled to ${\bf J}$ 
by taking the derivatives of the thermodynamic potential with respect to ${\bf J}$.
One example is the chiral condensate. 
Treating the current quark masses as external sources, 
one can use the Hellmann-Feynman theorem to calculate the chiral condensates
\begin{align}
\langle\bar{q}q\rangle=\frac{\partial\Omega}{\partial m_q}\,.
\end{align}
In practice, we construct the generating functionals for hadronic and quark matter 
and calculate the corresponding boundary conditions at $2n_0$ and $5n_0$.
These boundary conditions constrain the possible functionals in the interpolation region. 
In this way, the condensates in the interpolated domain capture the trends of both hadronic and quark matter.

\begin{figure*}\centering
\includegraphics[width=0.4\hsize]{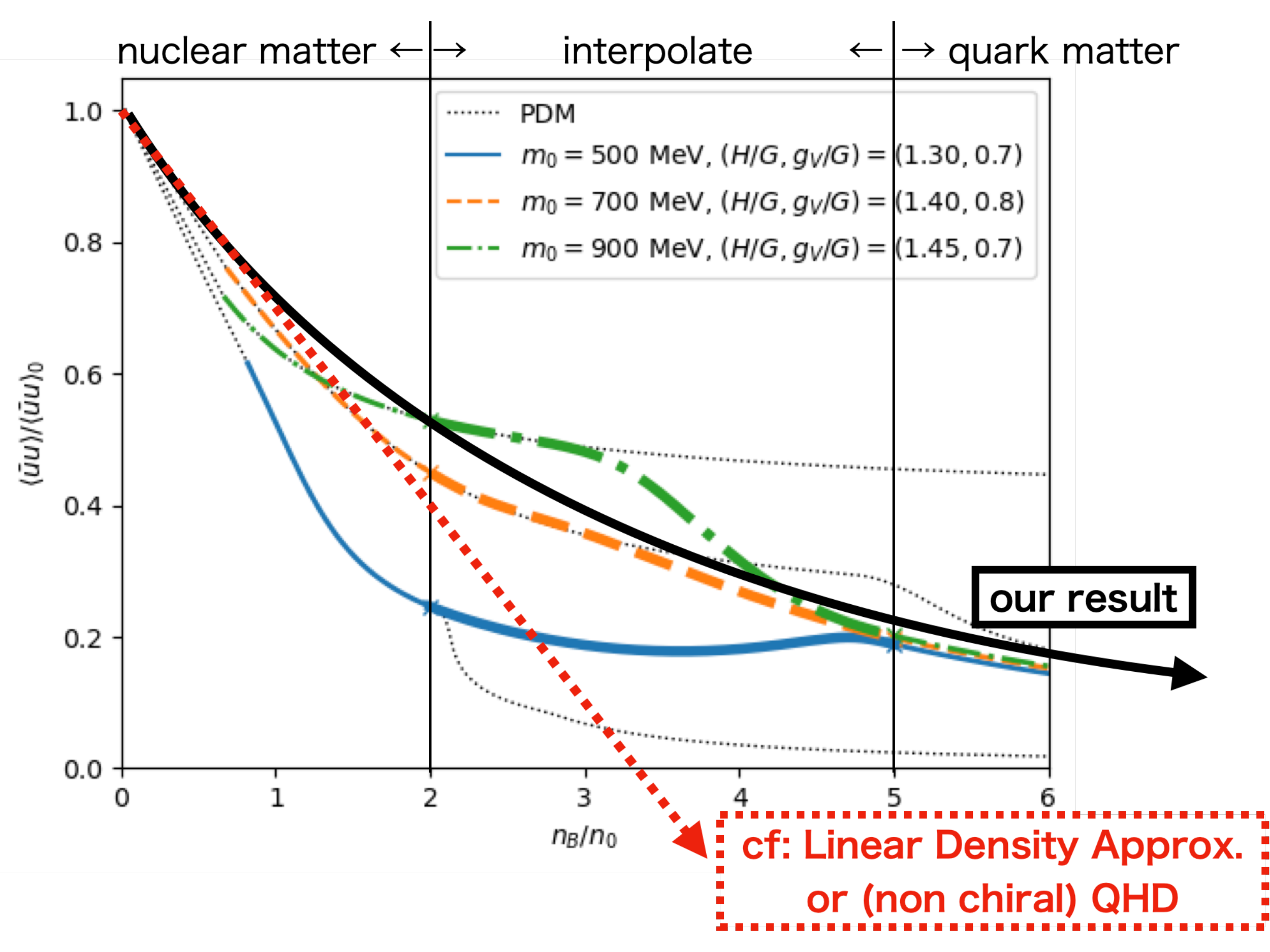}
\caption{
Ratio of the obtained chiral condensates to the vacuum values
vs. the baryon number density. 
All curves of our results remain positive and approach to zero gradually, 
as indicated by solid black arrow. 
On the other hand, 
the typical results obtained from e.g., linear density approximation or non-chiral QHD flip their signs 
at some high density as indicated by dotted red arrow. }
\label{fig-interpolated_chiral_condensate}
\end{figure*}

In the PDM, the trend of the chiral restoration is similar to what we infer from the linear density approximation which should be valid in dilute regime. 
The key observation is that nucleons carry the positive chiral scalar charges whose signs are opposite to the negative vacuum scalar density.
Hence increasing baryon density neutralizes the chiral scalar density $\sigma$.
Meanwhile, the extrapolation of this picture toward high baryon density eventually leads to the system with full of positive scalar charges, see Fig.3 for the flip of the sign in $\sigma$.
This seems odd with quark model predictions in which chiral condensates lose the magnitudes at high density; for instance, the NJL model predicts that the in-medium chiral condensates are about 20\% of the vacuum counterparts.
For the interpolated domain, these trends of hadronic and quark model descriptions are balanced by matching the interpolating functions with the hadronic and quark matter boundaries.
Figure \ref{fig-interpolated_chiral_condensate} is 
the ratio of the obtained chiral condensates to the vacuum values for various choices of $m_0$ and $(g_V,H)$. 
For comparison, we also show the extrapolation of the PDM and linear density approximation beyond the hadronic domain. 
We note that, while the PDM shows wild variations for different values of $m_0$ for the domain $n_B=2$-$5n_0$, our unified modeling tempers the $m_0$ dependence by putting quark model constraints from the high density side.

In qualitative terms, the hadronic and quark models 
are supposed to cover the different aspects of the chiral restoration.
In typical hadronic models, 
one does not manifestly consider the structural changes of nucleons around $n_0$.
Nevertheless the chiral restoration can take place simply due to the scalar charges inherent to nucleons.
In this picture, the reduction of spatial average of the vacuum and nucleon scalar densities does not immediately mean the significant reduction of the magnitude in vacuum and in nucleons for each.
With this description, the dynamical mass, e.g., constituent quark masses in a nucleon can remain large even after $\sigma$ is reduced by $\sim30$\% at $n_0$ and $\sim 40$-$80$\% at $2n_0$ as predicted by the PDM.
In the PDM, the chiral invariant mass $m_0$ plays the role to keep nucleons massive in hadronic domain.
In contrast, the quark models deal with the reduction of chiral condensates in magnitude not at the level of spatial average but at each location.
For instance the NJL model predicts the modification of the quark Dirac sea and associated reduction of the constituent quark masses. 
In terms of the chiral invariant mass $m_0$, the quark models at high density predict the reduction of $m_0$. 
Our interpolation balances these different chiral restoration in a somewhat balanced way.

%%%%%%%%%%%%%%%%%%%%%%%%%%%%%%%%%%%%%%%%%%%%%%%%%%%%%%%%%%%%%%%%
%	SUMMARY AND DISCUSSION
%%%%%%%%%%%%%%%%%%%%%%%%%%%%%%%%%%%%%%%%%%%%%%%%%%%%%%%%%%%%%%%%

\section{Summary and discussion}

%%% summary and comments on the results 

% result 1
Firstly, we showed that the PDM with the substantial chiral invariant mass, 
$600\,\mathrm{MeV}\lesssim m_0\lesssim900\,\mathrm{MeV}$, 
can provide $M$-$R$ relation of neutron stars consistently the observations.
In the future, the further observations of neutron stars may 
restrict the chiral invariant mass more strictly, 
because the typical radius of a neutron star (with e.g. $M\approx1.4M_\odot$) 
is strongly correlated with the property of the hadronic matter. 
Moreover, since the model parameters of hadronic and quark matter constrain each other 
through the thermodynamically consistent interpolation, 
the parameters of the NJL model can also be restricted at the same time. 

% result 2
Second, we showed that 
the chiral condensate in the hadronic matter 
where the change of the chiral condensate is driven by the scalar density of baryons, 
is smoothly connected to the one in the quark matter 
where the change is by the modification of quark Dirac sea.
%
%Our obtained chiral condensate tells 
%that the interpolation is a bridge between two different physics. 
PDM reflects the property that the nucleon mass does not change much, 
and this property is consistent with ignoring the density dependence of Dirac sea in the hadronic matter. 
On the other hand, Dirac sea plays an important role in the NJL model. 
It can be said that our crossover description connects the property of the hadronic matter and the change of the vacuum structure in the quark matter smoothly. 
This is one of the results that reflect the hadron-quark crossover picture.

%%% discussion about inhomo chiral cond 

Here, let us discuss the inhomogeneous chiral condensates qualitatively. 
A naive extrapolation of the linear density approximation indicates that 
the scalar charge of a nucleon almost cancels 
the chiral condensate in vacuum, so that 
the chiral condensate in a nucleon may have the opposite sign to the vacuum one. 
Therefore, when the baryon density increases, the space average of the chiral condensate can flip its sign. 
However, according to our result, the chiral condensate keeps its sign and approaches zero gradually. 
If the chiral condensate has a wave form in the sufficiently high density medium, 
the space average can vanish. 
%{\color{red}{ [! How about deleting the following sentence?  We may need some references when we write the following. !] }}
%This scenario is consistent with not only our results 
%but also some other results about inhomogeneous chiral condensates, 
%such as the dual chiral density wave or the Skyrme model. 

%%% future prospects

In this work, we assumed that the hyperons do not contribute in the lower density $n_B<2n_0$. 
If one extend the PDM to a three-flavor model and adopt its EOS up to even higher density e.g. $n_B<3n_0$, 
we can discuss the behavior of the strange quark contribution in neutron star more precisely.

\paragraph{Acknowledgments:}
The work of T.M. and M.H. was supported in part by JSPS KAKENHI Grant No. 20K03927. 
T.M. was also supported in part by the Department of Physics, Nagoya University. 
T.K. was supported by NSFC Grant No. 11875144.

%%%%%%%%%%%%%%%%%%%%%%%%%%%%%%%%%%%%%%%%%%%%%%%%%%%%%%%%%%%%%%%%
%%%%%%%%%%%%%%%%%%%%%%%%%%%%%%%%%%%%%%%%%%%%%%%%%%%%%%%%%%%%%%%%

\end{multicols}
\medline
\begin{multicols}{2}

\end{multicols}
\end{document}